\documentclass[prl,reprint,showpacs,amsmath,amssymb]{revtex4-1}

\usepackage{graphicx}
\usepackage{dcolumn}
\usepackage{bm}

\begin{document}

\title{Advantages of Mass-Imbalanced Ultracold Fermionic Mixtures
for Approaching Quantum Magnetism in Optical Lattices}

\author{Andrii Sotnikov}
\author{Daniel Cocks}%
\author{Walter Hofstetter}%
\affiliation{%
Institut f\"ur Theoretische Physik, Goethe-Universit\"at, 60438 Frankfurt/Main, Germany
}%

\date{\today}

\begin{abstract}
We study magnetic phases of two-component mixtures of ultracold fermions with repulsive interactions in optical lattices in the presence of hopping imbalance. Our analysis is based on dynamical mean-field theory (DMFT) and its real-space generalization at finite temperature. We study the temperature dependence of the transition into the ordered state as a function of the interaction strength and the imbalance parameter in two and three spatial dimensions. We show that below the critical temperature for N\'{e}el order mass-imbalanced mixtures also exhibit a charge-density wave, which provides a directly observable signature of the ordered state. For the trapped system, we compare our results obtained by real-space DMFT to a local-density approximation. We calculate the entropy for a wide range of parameters and identify regions, in which mass-imbalanced mixtures could have clear advantages over balanced ones for the purpose of obtaining and detecting quantum magnetism. 
\end{abstract}

\maketitle

Ultracold atoms in optical lattices have become in recent years very convenient and powerful systems for quantum simulations of condensed matter materials \cite{Ess2010ARCMP,Hof2002PRL,Blo2008RMP}.
In a variety of cases, these gases show even richer physics than solid state systems.
While spectacular progress has been made in realizing different many-body phenomena in the past few years, many quantum phases present in condensed matter physics still represent very challenging goals for cold gas experiments. 
One of these challenges is to obtain quantum magnetism in optical lattices.
Considerable progress towards this goal was made in a recent experiment~\cite{Jor2010PRL} with ultracold $^{40}$K prepared in a mixture of two different hyperfine states.

Here we focus on two-component mixtures of repulsively interacting fermions with an imbalance in the hopping amplitudes that can be realized, for example, by loading ultracold $^6$Li and $^{40}$K \cite{Tag2008PRL} or different alkaline earth atoms \cite{Tai2010PRL} into an optical lattice, or by using spin-dependent optical lattices for two different hyperfine states \cite{Man2003PRL}. 
We consider a Fermi-Hubbard Hamiltonian of the following type:
\begin{eqnarray}
\mathcal{\hat{H}}=&&
-t_A\sum\limits_{\langle i,j\rangle}(\hat{a}^\dag_{i}\hat{a}_{j}+{\rm H.c.})
-t_B\sum\limits_{\langle i,j\rangle}(\hat{b}^\dag_{i}\hat{b}_{j}+{\rm H.c.})
\nonumber\\
&&+U\sum\limits_{i}\hat{n}_{iA}\hat{n}_{iB}
+\sum\limits_{i}\sum\limits_{\alpha=A,B}(V_i-\mu_\alpha)\hat{n}_{i\alpha},
\label{eq.1}
\end{eqnarray}
where $t_A$ and $t_B$ are the hopping amplitudes of fermionic species $A$ and $B$, respectively.
%
The notation $\langle i,j\rangle$ indicates a summation over nearest-neighbor sites, and $U$ is the amplitude of the on-site repulsive ($U>0$) interaction of different species with corresponding densities $\hat{n}_{iA}$ and $\hat{n}_{iB}$.
In the last term, $V_i$ is the amplitude of the external (e.g., harmonic) potential at lattice site $i$, and $\mu_\alpha$ is the chemical potential of species $\alpha$ (henceforth we restrict ourselves to the case $\mu_A=\mu_B=\mu$).
The Hamiltonian~(\ref{eq.1}) corresponds to the single-band approximation; in other words, we consider the case of a sufficiently strong lattice potential, $V_{\text{lat}}\gtrsim 5E_r$. 
Thus, the hopping amplitudes $t_\alpha$ are given by
$
 t_\alpha \approx \frac{4}{\sqrt\pi}E_{r\alpha}^{1/4}{V_{\text{lat}}^{(\alpha)}}^{3/4}
 \exp\left(-2\sqrt{{V_{\text{lat}}^{(\alpha)}}/{E_{r\alpha}}}\right)
$
\cite{Zwe2003JOB},
where $E_{r\alpha}={\hbar^2k^2}/{2m_\alpha}$ is the recoil energy, $k$ is the wave number determined by the wavelength of the laser forming the optical lattice, and $m_\alpha$ is the mass of species $\alpha$.
The amplitude $V_{\text{lat}}^{(\alpha)}$ of the lattice potential can be different for the two components $V_{\text{lat}}^{(A)}\neq V_{\text{lat}}^{(B)}$, which results in the possibility to realize an imbalance in the hopping amplitude even for different hyperfine states of same atom ($m_A=m_B$) \cite{Man2003PRL}.

By using the Schrieffer-Wolff transformation in the limit $t_{A,B}\ll U$ near half filling, $n_{Ai}+n_{Bi}\approx1$, the Hamiltonian~(\ref{eq.1}) can be mapped onto an effective spin Hamiltonian \cite{Kuk2003PRL,Alt2003NJP}.
For the system under study, this transformation results in the anisotropic Heisenberg model,
\begin{equation}\label{eq.3}
  \mathcal{\hat{H}}_{\textrm{eff}}=J_{\parallel}\sum_{\langle ij\rangle}\hat{S}_{i}^{Z}\hat{S}_{j}^{Z}
 +J_{\perp}\sum_{\langle ij\rangle}(\hat{S}_{i}^{X}\hat{S}_{j}^{X}+\hat{S}_{i}^{Y}\hat{S}_{j}^{Y}),
\end{equation}
with coupling constants ${J_{\perp}=4t_At_B/U}$ and ${J_{\parallel}= 2(t_A^2+t_B^2)/U}$.
Note that here and below, all ``magnetic'' characteristics refer to a pseudospin made of two different species ($A$ and $B$), which corresponds to the spin operators $\hat{S}_{i}^{Z}=(\hat{a}^\dag_{i}\hat{a}_{i}-\hat{b}^\dag_{i}\hat{b}_{i})/2$ and $(\hat{S}_{i}^{X}+i\hat{S}_{i}^{Y})=\hat{a}^\dag_{i}\hat{b}_{i}$.
Thus it should be mentioned that the anisotropy in the spin model is not related to the actual spatial directions or quantization axes in the original optical lattice.

In the presence of hopping imbalance, i.e., $t_A\neq t_B$, the antiferromagnetic coupling $J_{\parallel}$ (in $Z$ direction) is always larger than the coupling $J_{\perp}$ (in the $XY$ plane). 
Note that in the limit of large hopping imbalance, $|t_A-t_B|\rightarrow(t_A+t_B)$, we find high anisotropy, $J_{\parallel}\gg J_{\perp}$; thus, the second term in the Hamiltonian~(\ref{eq.3}) can be considered as a perturbation, and we arrive at the Ising model.
Therefore, the imbalance in the hopping amplitude can be treated as a parameter that effectively reduces quantum fluctuations of the system under study.
As shown below, this reduction results, in particular, in a relative increase of the N\'{e}el-ordered region even for larger tunneling $t_\alpha\sim U$.

%

Note that the above mapping to the effective spin model is presented in order to give a better physical understanding of the system under consideration.
Naturally, for quantitative theoretical predictions in the intermediate coupling region ($t_\alpha\sim U$) it is necessary to use nonperturbative numerical methods, in combination with the full Hamiltonian~(\ref{eq.1}).
Below we apply dynamical mean-field theory (DMFT) \cite{Geo1996RMP} to examine the influence of hopping imbalance on the ordered phases in mixtures of ultracold repulsive fermions.

Before diving into details, let us introduce parameters that are useful for further analysis: the average hopping $t=(t_A+t_B)/2$ and the dimensionless hopping-imbalance parameter $\Delta t=(t_A-t_B) /(t_A+t_B)$. 
Unless otherwise specified, we consider atoms of type $A$ to be lighter than atoms of type $B$ ($t_A\geq t_B$), such that the hopping-imbalance parameter is positive, $\Delta t\in[0,1]$.

To study bipartite order, such as antiferromagnetism, it is necessary to consider a two-sublattice configuration within the self-consistency conditions of DMFT in the homogeneous case \cite{[{This configuration was also used in DMFT studies of two-component mass-imbalanced fermionic mixtures with attractive interactions ($U<0$) in }] Dao2007PRB}.
Note that physical observables corresponding to these sublattices, in contrast to the case of a balanced mixture ($t_A=t_B=t$), cannot be directly associated with quantities corresponding to the two different species. 
Instead, the Green's functions are defined as
$
{G}_{s\alpha}(i\omega_n) = \zeta_{\bar{s}\alpha} 
\int {D(\epsilon)}/({\zeta_{1\alpha} \zeta_{2\alpha} - \epsilon^2}) d\epsilon
$
\cite{Geo1996RMP},
where $s=\{1,2\}$ and $\bar{s}$ are the sublattice index and its opposite, respectively, 
$\zeta_{s\alpha} = i\omega_n + \mu_\alpha - \Sigma_{s\alpha}(i\omega_n)$, 
the Matsubara frequency corresponding to the temperature $T$ (we set $k_B=1$) is $\omega_n=\pi(2n+1)T$ with $n\in \mathbb{Z}$,
the self-energy $\Sigma_{s\alpha}(i\omega_n)$ carries information about local interactions, 
and $D(\epsilon)$ is the density of states [we present results for $D(\epsilon)$ corresponding to cubic or square lattice geometries below].
To solve the impurity problem, we use the exact diagonalization solver \cite{Geo1996RMP}.
We also verify our results for selected parameters by performing calculations with other highly accurate numerical algorithms, such as the continuous-time Monte Carlo \cite{Gul2011RMP} and numerical renormalization group \cite{Bul2008RMP} impurity solvers.

\begin{figure}
\includegraphics{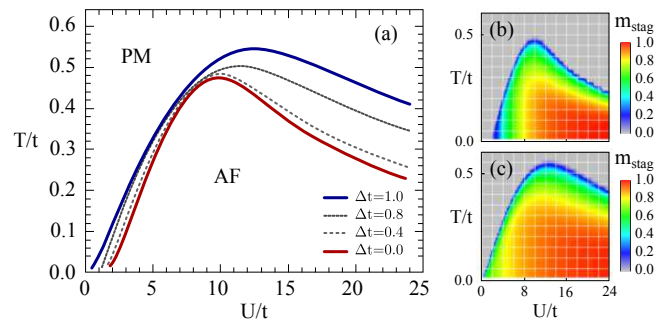}
    \caption{\label{01TUdiag}
    (a) Critical temperature for N\'{e}el ordering of the half filled Hubbard model at different values of the mass-imbalance parameter in a cubic lattice, obtained within DMFT. (b), (c) Contour plots indicating the absolute value of the order parameter $m_{\text{stag}}=|n_{Ai}-n_{Bi}|$ for the balanced (b), $\Delta t=0$, and maximally-imbalanced (c), $\Delta t=1.0$, mixtures.}
\end{figure}

For the homogeneous system, we concentrate on the case of half filling, $\mu=U/2$. 
At low temperatures its ground state exhibits N\'{e}el (antiferromagnetic) ordering.
The corresponding phase diagram obtained by DMFT for the cubic lattice is depicted in Fig.~\ref{01TUdiag}.
One can conclude that the hopping imbalance results in a relative increase of the critical temperature (note that it is scaled in units of average hopping $t$).
In the atomic limit ($U\gg t$), one can derive this from the mean-field value of the N\'{e}el temperature in the Heisenberg model, 
$T_{N} ={6JS(S+1)}/{3}$ \cite{Ashcroft1976}, where $S$ is the fermions' spin. 
Taking $J=J_{\parallel}=2(t_A^2+t_B^2)/U$, one obtains for constant ${U}/{t}$ the relation ${T_{N}(\Delta t)}/{T_{N}(0)} = 1+\Delta t^2$.

We also perform analogous calculations for a two-dimensional system with square lattice geometry. 
%
We would like to emphasize here that the Mermin-Wagner theorem, which excludes long-range order at finite temperature in low dimensions ($d\leq2$) for continuous symmetries, 
is applicable only to the rotation symmetry of the Hamiltonian~(\ref{eq.3}) in the $XY$ plane. Thus, for $J_{\parallel}>J_{\perp}$, long range $Z$-antiferromagnetic order is permitted \cite{Ashcroft1976}.
One can verify that, even for $d=1$ \cite{Caz2005PRL}, the excitation (magnon) spectrum has an energy gap $\Delta\propto(J_{\parallel} - J_{\perp})$. This protects the system from the infinite number of low-energy excitations, which destroy antiferromagnetic order in the balanced ($t_A=t_B$) system at $T>0$ in $d\leq2$.
Regardless of whether hopping imbalance is present or not, a critical temperature in $d=2$ can be defined if one uses a mean-field approach (or DMFT, as applied here).
Our results for the dependence of the transition temperature on the imbalance parameter and interaction strength are similar to the dependence for that of the cubic lattice as shown in Fig.~\ref{01TUdiag} but with lower values of the critical temperature; e.g., we obtain $\max[T_{c}^{2d}(\Delta t=0)]\approx0.39t$ and $\max[T_{c}^{2d}(\Delta t=1)]\approx0.45t$.

It should be noted that the data obtained for the transition temperatures in the limiting case of a mixture with equal hopping amplitudes ($\Delta t=0$) quantitatively agree with results presented in Refs.~\cite{Wer2005PRL, Gor2010PRL, *Gor2011JLTP}, where DMFT was also used.
We emphasize that it has been well-established that, in the balanced case, DMFT overestimates the critical temperature for moderate to large interactions ($U/t>5$) in comparison with exact quantum Monte Carlo (QMC) simulations \cite{Staudt2000EPJB,Fuchs2011PRL} and a dynamical cluster approximation (for quantitative comparison see \cite{Kent2005PRB}).
However, in the limit of large interactions, existing calculations
\footnote{Exact solutions of the Heisengerg \cite{Sandvik1998PRL} and Ising  \cite{Talapov1996JPA} models that correspond in the limit $U\gg t$ to the cases $\Delta t=0$ and $\Delta t=1$, respectively, show a considerable increase of the critical temperature $T_{c}^{I{\rm (ex)}}/T_{c}^{H{\rm (ex)}}\approx2.4$, which is even larger than $T_{c}^{I{\rm (mf)}}/T_{c}^{H{\rm (mf)}}=2$.}
show that the qualitative increase of $T_{c}$ due to an imbalanced mixture will in fact be enhanced by the use of an exact method such as QMC calculations. We expect that this is also true for intermediate interactions.

\begin{figure}
\includegraphics{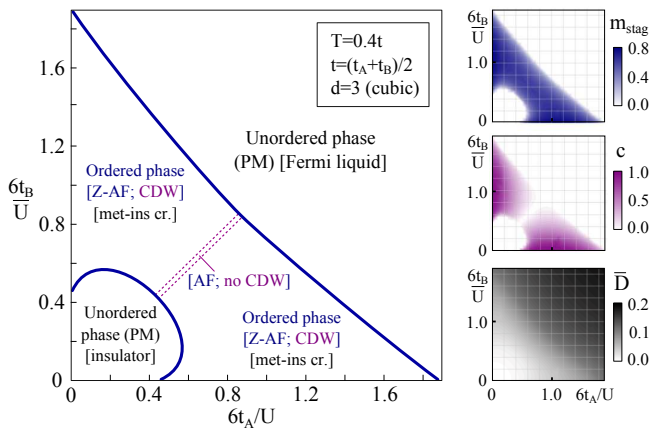}
    \caption{\label{02tatb}
    Phase diagram of a two-component fermionic mixture obtained by DMFT at finite temperature. 
    The temperature is scaled proportionally to the average hopping, $T=0.4t$.
    The contour plots to the right indicate the absolute value of the staggered magnetization $m_{\text{stag}}$, the charge-density wave parameter $c=\overline{|D-K|}/\overline{(D+K)}$, and the average double occupancy $\overline{D}$, where the average is taken over the sublattices $s=1,2$.
    }
\end{figure}

One can also gain important information from a phase diagram, where the hopping amplitudes of different species are scaled independently (for bosons, see Refs.~\cite{Hub2009PRB,Soy2009NJP}), which is presented in Fig.~\ref{02tatb}.
It contains information on both magnetic ordering and the conductive (localization) properties of the system. 
The N\'{e}el phase is characterized by an order parameter, the staggered magnetization, equivalent to a spin-density wave.
However, from Fig.~\ref{02tatb} it is clear that in the mass-imbalanced mixture one has an additional type of order, {\it a charge-density wave} (CDW), which, in general, is nonzero inside the ordered phase except for equal hopping amplitudes, $t_A=t_B$ (balanced mixture), where it vanishes.
To define whether the system is in the metallic or insulating state at half filling, we also analyze both the double occupancy, $D_i=\langle \hat{n}_{Ai} \hat{n}_{Bi} \rangle$, and the zero occupancy, $K_i=\langle (1-\hat{n}_{Ai}) (1-\hat{n}_{Bi}) \rangle$.

\begin{figure}
\includegraphics{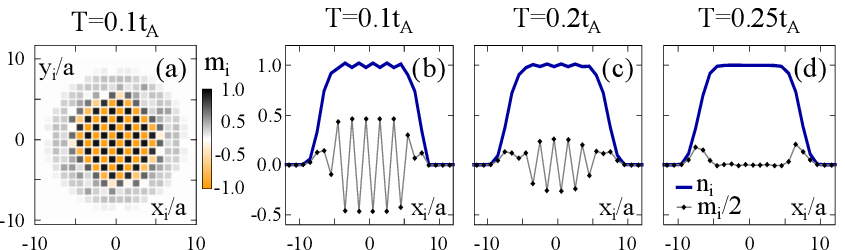}\\[2mm]
\includegraphics{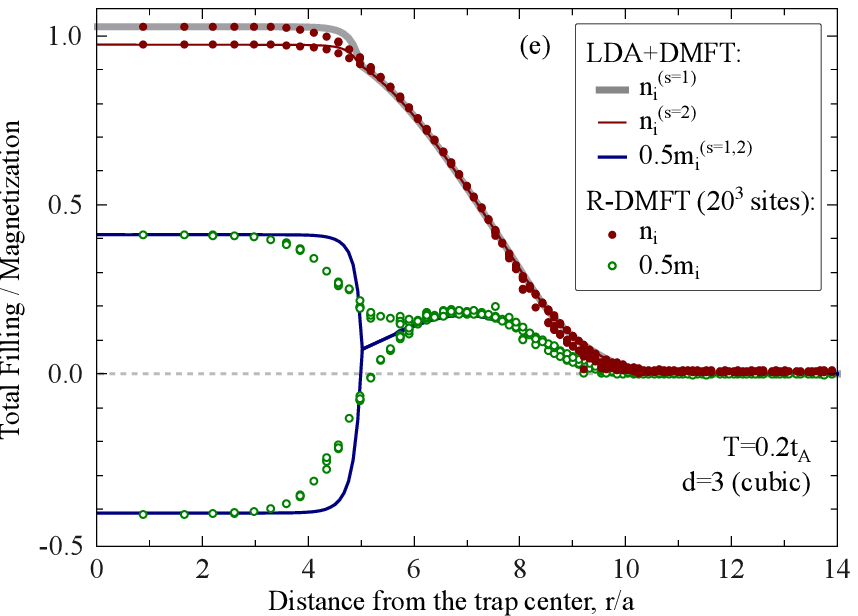}
    \caption{\label{03rdmft}
    Upper row (a)-(d): Magnetization and total density distribution of mass-imbalanced repulsive fermions in a harmonic trap at different temperatures (R-DMFT, square lattice, $d=2$).
    (e) The same distributions obtained within the local-density approximation (LDA+DMFT, lines) and R-DMFT (dots, corresponding to individual sites) at temperature $T=0.2t_A$ for a cubic lattice in $d=3$. 
    Other parameters used in (a)-(e): $t_B=0.5t_A$, $U=10t_A$, $\mu=U/2$, and $V_i=0.1t_A(r_i/a)^2$, where $a$ is the lattice constant.
    }
\end{figure}

The origin of the charge-density wave can be more clearly understood from the analysis of spatial distributions. 
To this end, we apply an extension of DMFT, {\it the real-space dynamical mean-field theory} (R-DMFT) \cite{Hel2008PRL,Sno2008NJP}.
It effectively captures long-range order together with effects originating from a finite size of the system and an external trapping (harmonic) potential. 
In Fig.~\ref{03rdmft} 
we show real-space distributions of the total filling $n_i=n_{Ai}+n_{Bi}$ and the occupation difference (magnetization) $m_i=n_{Ai}-n_{Bi}$ both in two- and three-dimensional optical lattice geometries.

From Figs.~\ref{03rdmft}(b)-(d) we see that the emergence of a charge-density wave at half filling ($n_i\approx1$) corresponds to the fact that in the N\'{e}el-ordered state the sites $i$, which are occupied by the heavier component, have a larger total filling ($n_i>1$); i.e., they have an enhanced double occupancy $D_i$ ($D_i>K_i$).
At the same time, the sites $j$ that are mainly occupied by the lighter component have a smaller total filling ($n_j<1$); i.e., they have an enhanced zero occupancy $K_j$ ($K_j>D_j$) due to a higher mobility of the lighter atoms.
One could possibly use this property as an alternative signature of magnetic ordering of ultracold hopping-imbalanced fermions in optical lattices, e.g., by performing site-resolved detection of atom pairs.

It should be noted that in the real-space density distributions of mass-imbalanced mixtures with $\mu_A=\mu_B$ there is always a ``ferromagnetic'' ring present [it is directly visible in Fig.~\ref{03rdmft}(a)]. 
This ring originates from the larger kinetic energy of the lighter component that, in turn, results in a wider spatial distribution of these atoms in a trap. 
We conclude from Figs.~\ref{03rdmft}(b)-(d), that this ring remains present at temperatures above the critical temperature and, thus, cannot be uniquely identified with quantum magnetism.

In Fig.~\ref{03rdmft}(e), we present a comparison of two computational approaches: R-DMFT and DMFT combined with a local-density approximation (LDA). 
We find very good agreement in the center of the trap (including the magnitude of the charge-density wave) and at the edges.
But in the intermediate region between the antiferromagnetic core and the ``ferromagnetic'' ring, LDA fails to reproduce the detailed structure as it does not account for a {\it proximity effect}; i.e., it neglects the influence of surrounding sites with different densities.
In contrast, R-DMFT shows a wider spatial range of stability of the antiferromagnetic order that results in an interesting interplay between antiferromagnetic correlations and the ``ferromagnetic'' ring.
Note that proximity-induced antiferromagnetic order in a trapped system has also been shown previously in R-DMFT studies of balanced mixtures \cite{Hel2008PRL,Sno2008NJP}.

\begin{figure}
\includegraphics{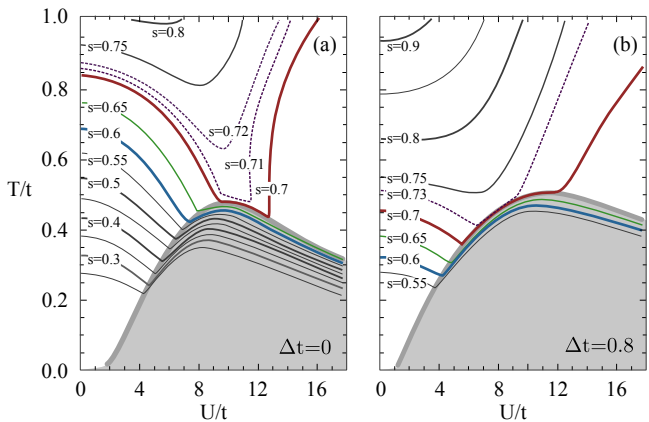}
    \caption{\label{04entr}
    Phase diagrams of the half filled Hubbard model ($\mu_0=U/2$) for balanced (a) and imbalanced (b) mixtures obtained within DMFT in the homogeneous (nontrapped) case. Shaded areas correspond to the N\'{e}el (antiferromagnetic) phase. The colored and dark-gray lines are isentropic curves.}
\end{figure}

Finally, as our approach allows for the extraction of expectation values of the local filling with high precision, one can also perform an entropy analysis for the homogeneous system.
Using a Maxwell relation for the entropy~$s$ per lattice site~$i$, $\partial s/\partial\mu = \partial n/\partial T$, where $n=\sum_{s}n^{(s)}_i/2$, one obtains
$
 s(\mu_0,T) = \int_{-\infty}^{\mu_0}({\partial n}/{\partial T})d\mu.
$
Therefore, by performing calculations at different values of the interaction strength $U$ and temperature $T$, we obtain the isentropic curves presented in Fig.~\ref{04entr}.

As one can conclude from Fig.~\ref{04entr}, the approach towards quantum magnetism in optical lattices based on preparation of the atomic mixture in the Fermi-liquid state, and consequent adiabatic increase of the interaction also works for the case of mass-imbalanced fermions.
The results obtained from our DMFT calculations show that, in order to cross the transition boundary, one needs to prepare the system in a state with entropy per particle $s<\ln2$ 
\footnote{DMFT overestimates the critical entropy value due to the neglect of spatial spin-spin correlations, see Refs.~\cite{Jor2010PRL, Wer2005PRL, Fuchs2011PRL, Leo2011PRA} for details}, independently of the value of the mass-imbalance parameter.
However, if one starts with higher entropy values, then it is more favorable to use imbalanced mixtures as they allow a much closer approach to the critical region by adiabatic increase of the interaction.
For example, at $s=0.75$ one obtains $T_{\min}=0.82t$ for $\Delta t=0$, while at the same entropy value $T_{\min}=0.51t$ for $\Delta t=0.8$.
One can also see that at the same relative values of the temperature in the Fermi-liquid region ($U\lesssim t$) the entropy increases with hopping imbalance.

It should be noted that Fig.~\ref{04entr} is obtained for the homogeneous (i.e., nontrapped) case.
Thus, the depicted isentropic curves include only one of the possible mechanisms of cooling, namely the Pomeranchuk effect in two-component ultracold fermionic mixtures \cite{Wer2005PRL}.
Let us emphasize that, in addition to this effect, the trapped system can also be cooled by a redistribution of atoms within the trap (see Refs.~\cite{Leo2011PRA,Pai2011PRL,Col2011NJP} for details).

To summarize, in this Letter we demonstrate advantages of hopping-imbalanced two-component mixtures of ultracold fermions for approaching quantum magnetism.
We compare critical temperatures and isentropic curves obtained by DMFT and show that for imbalanced mixtures the critical region can be approached at higher entropy values than for balanced mixtures.
We also observe that hopping-imbalanced ultracold Fermi gases in the N\'{e}el-ordered state have additional long-range order in the form of a charge-density wave, which provides an alternative signature for detecting quantum magnetic order in optical lattices.

\begin{acknowledgments}
We are thankful to L. He, M. Snoek, and I. Titvinidze for useful discussions.
Support by the German Science Foundation DFG via Forschergruppe FOR 801 and Sonderforschungsbereich SFB/TR 49 is gratefully acknowledged.
\end{acknowledgments}

\bibliography{A14}

\end{document}